\documentclass[osajnl,twocolumn,showpacs]{revtex4} 

\usepackage{epsfig,bm,times}

\begin{document}

\title{Direct measurement of spatial Wigner function with area-integrated detection}

\author{Eran Mukamel, Konrad Banaszek, and Ian A. Walmsley}

\affiliation{Clarendon Laboratory, University of Oxford, Parks Road,
Oxford OX1 3PU, United Kingdom}

\author{Christophe Dorrer}

\affiliation{Bell Laboratories, Lucent Technologies,
101 Crawfords Corner Road, Holmdel, NJ 07733}

\date{\today}

\begin{abstract}
We demonstrate experimentally a novel technique for characterizing transverse spatial
coherence using the Wigner distribution function. The presented method is based on measuring interference between a pair of rotated and displaced replicas of the input beam with an area-integrating detector, and it can be superior in regimes when array detectors are not available. We analyze the quantum optical picture of the presented measurement for single-photon signals and discuss possible applications in quantum information processing.
\end{abstract}


\maketitle

The problem of characterizing spatial coherence of optical fields appears in a variety of applications. In biomedical optics, spatial coherence is being studied as a means to reveal the internal structure of scattering media, thus providing novel modes of tomographic imaging.\cite{WaxYangJOSAA02} In optical communications, spatial coherence of light sources is necessary to achieve efficient coupling into fiber systems. This aspect becomes even more critical in quantum communications,\cite{MichKiraSCI00,KurtMayePRL00,BeveBrouPRL02} where non-classical sources of radiation are used, and detection is often based on subtle interference effects that require good mode matching.

In this paper we present a novel technique for characterizing spatial coherence of light in the Wigner representation.\cite{WignPhR32,BastOpC78,Dragoman} Our technique has several features that make it distinct from previously proposed schemes. First, it is based on a series of measurements performed with an area integrating detector that collects all available light power, rather than on spatially resolved detection employed is some of the previous proposals.\cite{BrenLohmOpC82,McAlBeckOpL95} Therefore, it can be used in regimes where array detectors are not available, for example in mid- and far infrared, or when single photon signals are involved. In such cases, our technique provides the optimal signal-to-noise ratio, analogously to the case of Fourier transform spectroscopy which is more efficient than a spectrometer with a single scanning detector. Compared to the recently proposed heterodyne schemes for measuring the Wigner function,\cite{WaxThomOpL96,LeeReilOpL99} our method is self-referencing and it does not require any auxiliary sources of radiation.

Our technique is based on the following observation.
For a complex quasi-monochromatic field $E({\bm\xi})$ varying in the transverse plane parameterized with the two-dimensional vector $\bm\xi$, the Wigner function is defined according to\cite{BastOpC78,Dragoman}
\begin{equation}
\label{Eq:WignerDefinition}
W({\bm x},{\bm k})=\frac{1}{\pi^2} \int d^2{\bm{\xi}} e^{2ik{\bm\xi}}
\langle E^{\ast}({\bm x}-{\bm\xi}) E({\bm x}+{\bm\xi}) \rangle.
\end{equation}
Here $\bm k$ is the transverse spatial frequency vector and the
angular brackets denote a statistical average. The above definition can be easily transformed to the form:
\begin{equation}
\label{Eq:Parity}
W({\bm x},{\bm k}) = \frac{1}{\pi^2} \left\langle \int d^2{\bm{\xi}}
[e^{-i{\bm k}{\bm\xi}}E({\bm x}-{\bm\xi})]^\ast
e^{i{\bm k}{\bm\xi}}E({\bm x}+{\bm\xi})
\right\rangle.
\end{equation}
This expression shows that the Wigner function at a point $({\bm x},{\bm k})$ is given
by a spatially integrated overlap of the field $e^{i{\bm k}{\bm\xi}}E({\bm x}+{\bm\xi})$ with a complex conjugate of its replica rotated by $180^0$, which corresponds to the transformation ${\bm\xi} \rightarrow - {\bm\xi}$. The field $e^{ik{\bm\xi}}E({\bm x}+{\bm\xi})$ can be obtained from the original input by two simple transformation: displacement in space by ${\bm x}$ and changing the direction of propagation by ${\bm k}$, which in the paraxial approximation results in multiplying the field by a phase factor $e^{ik{\bm\xi}}$.

\begin{figure}[b]
\centerline{\epsfig{file=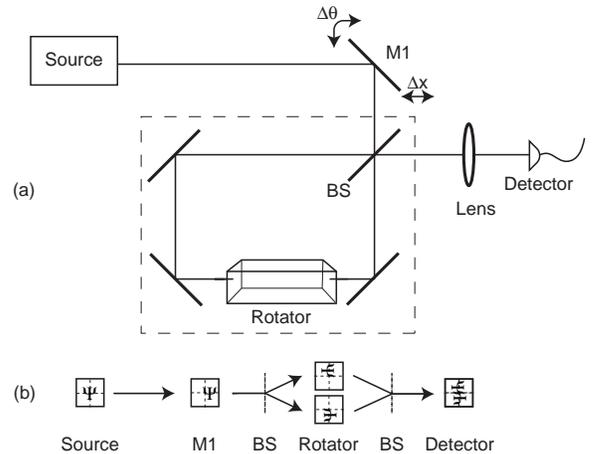,width=3in}}
\caption{Experimental apparatus for measuring the Wigner function.\label{Fig:Setup}}
\end{figure}

The above idea can be implemented to measure the Wigner function in a setup depicted in Fig.~\ref{Fig:Setup}(a). The input beam is steered using the mirror M1 into a three-mirror Sagnac interferometer. The displacement and the tilt of the mirror control the point at which the Wigner function is measured.
The 50:50 beam splitter BS generates a pair of replicas of the displaced and tilted input beam travelling in the opposite directions. The interferometer contains a Dove prism whose base forms $45^0$ with the plane of the interferometer. Transmission through the Dove prism rotates each of the counterpropagating beams by $90^0$, which adds up to the $180^0$ relative rotation required to measure the Wigner function. The operation of the setup can be most easily understood by analyzing propagation of the replicas of an exemplary image through the interferometer, as shown in Fig.~\ref{Fig:Setup}(b). The counterpropagating replicas of the input field are recombined at the output port of the interferometer, and the emerging beam is focused
with the help of a lens on an area-integrating detector.

The intensity $I$ recorded by the detector 
can be decomposed into a sum of three terms, $I = I_1 + I_2 + I_{12}$.
The first two of them are given by the space-integrated input intensity $|E(\bm{\xi})|^2$ and they remain constant as long as the aperture of the interferometer does not clip off any of the input field.
The third term $I_{12}$, originating from the interference between the counterpropagating replicas is proportional to the right hand side of Eq.~(\ref{Eq:Parity}):
\begin{equation}
I_{12} = \frac{\pi^2\cos \varphi}{2}  W({\bm x}, {\bm k})
\end{equation}
thus giving the value of the Wigner function. 
Here $\varphi$ is the relative phase between the interfering replicas.
The complete Wigner function can be therefore scanned by measuring the detector photocurrent as a function of the position and the tilt of the steering mirror, and subtracting the constant pedestal. We note that the factor relating the interference term to the Wigner function depends on the relative phase $\varphi$ between the interfering fields, and it can assume in principle either a positive or negative sign. We also note that if we extract the beam leaving through the second port of the interferometer, for example with the help of an optical circulator, then the constant pedestal can be removed by subtracting photocurrents measured at both the output ports of the interferometer.

We used the presented technique to measure the Wigner function for several beams derived from a He-Ne laser. For simplicity, we restrict ourselves only to one spatial dimension, parallel to the plane of the interferometer. The transverse distributions of the electric field for the prepared beams were factorizable in the directions parallel and perpendicular to the interferometer plane, so that the input field can be effectively considered as one dimensional. The steering mirror was mounted on a pair of translation and rotary stages (Physik Instrumente, models M--014.D0 and M--036.D0) controlled by a computer, which also recorded the value of the photodiode current. The interferometer was constructed using 1 inch diameter mirrors and a beam splitter, and a Dove prism 15~mm in cross-section. The beam emerging from the interferometer was focused with an $f=25$~mm lens on a detector with active area 3.6~mm $\times$ 3.6~mm.

\begin{figure*}
\centerline{\epsfig{file=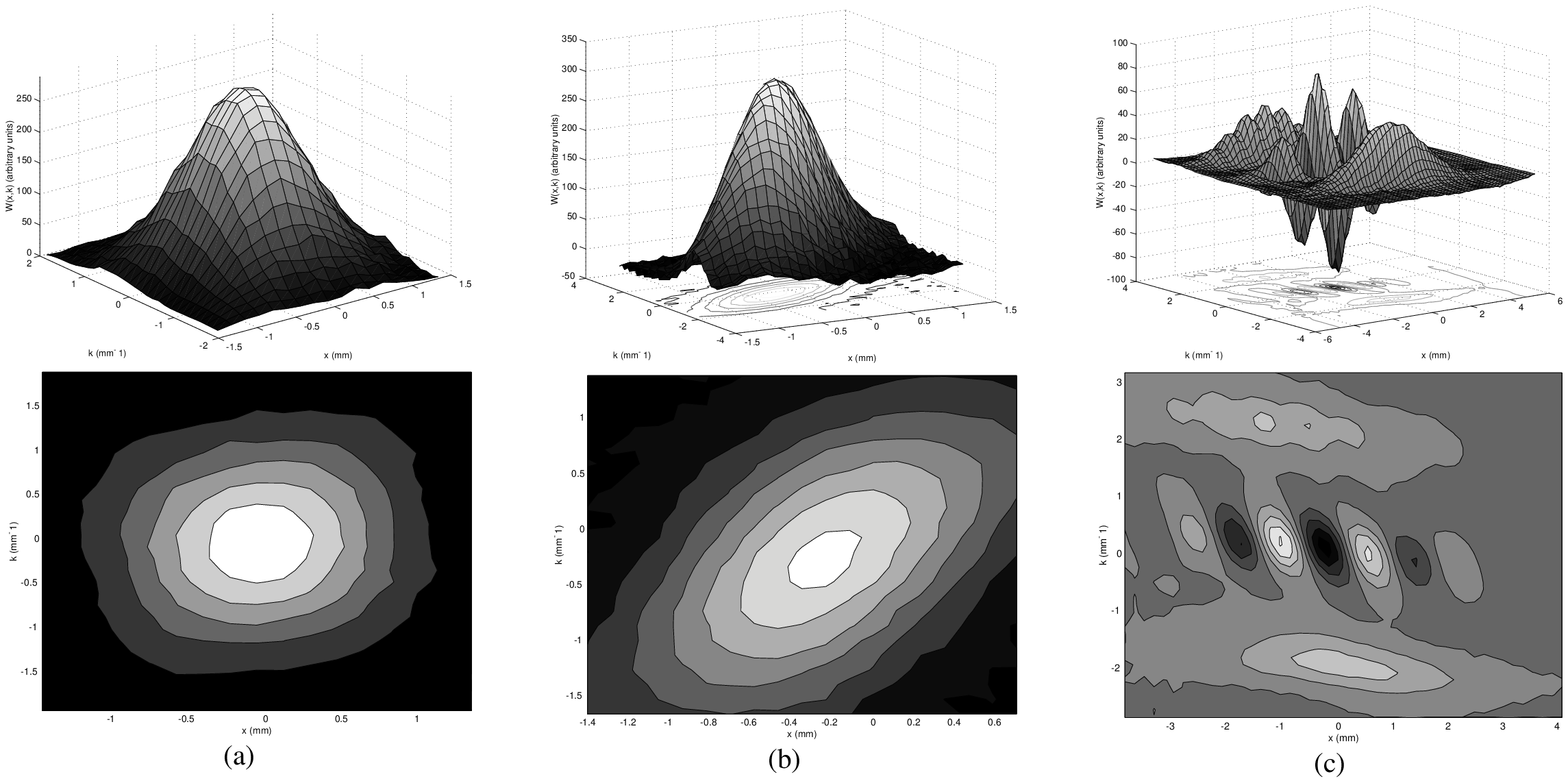,angle=90,width=7in,bburx=460,bbury=780,bbllx=125,bblly=110}}
\caption{Surface and countour plots of Wigner distribution functions measured for (a) collimated and (b) slightly divergent single mode beams, and (c) a laser beam reflected off a glass wedge.\label{Fig:Results}}
\end{figure*}

In Fig.~\ref{Fig:Results} we depict the measured Wigner functions. The first two plots were obtained for a laser beam filtered through a single-mode fiber and (a) collimated or (b) made lightly divergent. These data show Gaussian Wigner functions with diffraction-limited widths along the principal axes. The tilt of principle axis in Fig.~\ref{Fig:Results}(b)
reflects positive correlations between the position and the direction of propagation characteristic for a diverging beam. Fig.~\ref{Fig:Results}(c) shows the Wigner
function for an input obtained by a reflection from the front and the back surfaces of glass wedge. The measured Wigner distribution function shows two slightly converging Gaussian beams with different central transverse wave vectors. The oscillating pattern located between the Gaussians is a signature of their mutual coherence. We fitted our experimental data to a theoretical model of a Wigner function describing a pair of partially coherent Gaussian beams, which yielded the degree of coherence equal to $1.02 \pm 0.04$. An additional feature seen in Fig.~\ref{Fig:Results}(c) is the presence of ripples on one of the Gaussian peaks, at twice the spatial frequency of the fringe pattern at the center of the distribution. These ripples can be attributed to interference with a third, much weaker beam generated by a secondary reflection from the glass wedge.

In the theoretical analysis presented above, we assumed that the range of transverse spatial frequencies in the input beam is narrow enough to make diffraction effects across the setup negligible. In a geneneral case, the measured Wigner function describes the distribution of the electric field in the plane in which the two replicas experience relative rotation. It is therefore critical that the replicas counterpropagating in the Sagnac interferomer are inverted at the same distance from the beam splitter BS that generated them. Diffraction after the Dove prism is irrelevant, as it does not change the total intensity which is the only quantity measured on the output. Also, when diffraction cannot be neglected, the parameterization of the Wigner distribution function obtained from the displacement and the tilt of the steering mirror needs to be compensated for the propagation distance from the steering mirror to the inversion plane. The compensation is given by a simple linear transformation of the Wigner function variables.\cite{BastOpC78,Dragoman}

One important regime of optical measurements where array detectors are not easily available are single-photon signals. When the output of the interferometer is measured using a detector with single-photon sensitivity, for example an avalanche photodiode, the presented
setup can be used to characterize transverse coherence of the output produced by single photon sources.\cite{MichKiraSCI00,KurtMayePRL00} Development of such sources is presently a subject of extensive research effort owing to prospective applications in quantum cryptography.\cite{BeveBrouPRL02} As secure quantum communication is most likely to be based on single-mode fiber links, efficient matching of the modal structure of the source to the fiber is necessary to achieve high transmission rates and close options for eavesdropping attacks.

The Wigner distribution function defined in Eq.~(\ref{Eq:WignerDefinition}) characterizes second-order statistical properties of fluctuating classical electromagnetic fields. However, it admits also a quantum mechanical interpretation which has a number of interesting consequences for our work. If the input signal consists of a single quasimonochromatic photon, then $E({\bm\xi})$, when properly normalized, is the wave function of that photon for the transverse spatial degrees of freedom.\cite{IBB} This correspondence has been recently used to propose encoding of information into spatial modes of a single photon characterized by different orbital angular momenta.\cite{LeacPadgPRL02} Separation of these modes was performed by interference between mutually rotated replicas in a manner similar to that underlying our scheme. 

Following the quantum mechanical picture, the spatial degree of freedom of a photon is another optical realisation of a continuous-variable quantum system, alternative to the quadrature of a single light mode. 
There exists an analogy between the scheme proposed here and a direct measurement of the quadrature Wigner function of a light mode.\cite{WallVogePRA96,BanaRadzPRA99} 
In the latter scheme, the photon number parity operator is the equivalent of the spatial rotation between the interfering replicas that yields the Wigner function in the present case. This analogy has interesting consequences when generalized to composite systems. Consider a pair of photons with perfectly correlated positions, described by a pure entangled wave function. Such a pair exemplifies the Einstein-Podolsky-Rosen paradox, and it forms a source of entanglement. For example, it can be used to test Bell's inequalities by sending the two photons to a pair of Sagnac interferometers and measuring the joint spatial Wigner function at selected points of the phase space. Such a measurement reveals quantum nonlocality, analogously to the joint measurement of displaced parity operators on a pair of quadrature entangled beams that has been discussed previously.\cite{BanaWodkPRA98}

Concluding, we presented a method for measuring transverse spatial coherence of optical beams in the Wigner representation. The scheme is particularly beneficial in situations when array detectors are not available. It has a number of implications in quantum optics and quantum information processing: it can be used as a diagnostic tool for novel sources of nonclassical radiation, and also probe spatial entanglement in optical systems.
This research was supported by ARDA Quantum Computing Program.

\end{document}